\documentclass[showpacs,aps,prl,twocolumn]{revtex4}
\usepackage{dcolumn}
\usepackage{bm}
\usepackage{graphicx}
\usepackage{amsmath}
\usepackage{latexsym}
\usepackage{amsfonts}
\usepackage{amssymb}
\usepackage{epsfig}
\usepackage{array}

\newcommand{\ket}[1]{\left\vert#1\right\rangle}
\newcommand{\ketbra}[2]{|#1\rangle \langle#2|}

\begin{document} 

\title{Experimental Demonstration of Decoherence-Free One-Way Information Transfer}

\author{R. Prevedel$^1$, M. S. Tame$^2$, A. Stefanov$^{1,3}$, M. Paternostro$^2$} 
\author{M. S. Kim$^2$}
\author{A. Zeilinger$^{1,3}$}
\affiliation{$^1$Faculty of Physics, University of Vienna, Boltzmanngasse 5, A-1090 Vienna, Austria\\
$^2$School of Mathematics and Physics, The Queen's University, Belfast, BT7 1NN, UK\\
$^3$Institute for Quantum Optics and Quantum Information (IQOQI), Austrian Academy of Sciences, Boltzmanngasse 3, A-1090 Vienna, Austria}

\date{\today}

\begin{abstract}
We report the experimental demonstration of a one-way quantum protocol reliably operating in the presence of decoherence. Information is protected by designing an appropriate decoherence-free subspace for a cluster state resource. We demonstrate our scheme in an all-optical setup, encoding the information into the polarization states of four photons.
A measurement-based one-way information-transfer protocol is performed with the photons exposed to severe symmetric phase-damping noise. Remarkable protection of information is accomplished, delivering nearly ideal outcomes.
\end{abstract}


\maketitle

Decoherence, the ubiquitous loss of information encoded in a quantum system due to its uncontrollable interaction with an environment, is one of the main obstacles in the grounding of quantum technology for massively parallel information processing.  
The design and testing of fault-tolerant protocols counteracting the effects of decoherence is thus necessary for the achievement of reliable quantum information processing (QIP). In a coarse-grained but yet effective picture, there are two distinguished strategies to overcome this problem. The first is quantum error correction (QEC) and deals with the correction of errors by means of proper codewords~\cite{1}. 
The second approach, decoherence-free subspace's (DFS's), prevents or reduces the effects of a {\it dominant} decoherence mechanism~\cite{11} by using symmetries in the system-environment coupling. This involves searching for a region of the Hilbert space that is inaccessible to the particular environment of a given physical system, thus allowing information to be encoded and protected.
It is known that DFS's can be concatenated with QEC~\cite{18}, the quantum Zeno effect~\cite{19} and dynamical decoupling~\cite{n18b}. Experimentally, DFS's have been tested in linear optics~\cite{21, 22}
as well as other systems~\cite{24, n25}. Here, different to these earlier studies, 
we successfully demonstrate the first use of a DFS for achieving one-way QIP in a 
system subject to phase-damping (PD) decoherence.

The one-way model for quantum computation (QC)
has recently received much attention in virtue of 
many appealing features~\cite{2,9}. 
Theoretical and experimental efforts have culminated in the generation of photonic cluster states~\cite{4, 5, 7, 8, n8}, the demonstration of the basic aspects of the model~\cite{4}, simple quantum algorithms~\cite{7,8,n8,10}, the design of techniques for the growth of large clusters~\cite{27} and the experimental correction of the inherent randomness in the measurement outcomes~\cite{7}. Although a threshold for fault-tolerance has been estimated~\cite{29}, there has so far been no experimental demonstration of noise-resilient one-way QC. This is an important step toward the upgrading of the model as a viable route for scalable QIP. Here, we perform such a step by demonstrating the encoding of a photonic resource into a DFS cluster state of two qubits (two-level systems) and carrying out a basic protocol on it. This is significantly different to any previous demonstration of DFS's~\cite{21,22,24,n25}; Our experiment highlights the benefit of using DFS's for one-way QIP. The novel strategy we use is shown to be effective for manipulating quantum information in the presence of noise; the cluster is shielded from the action of symmetric PD decoherence mechanisms~\cite{1}. The model complements the use of QEC in a one-way scenario~\cite{29}. In our experiment we have demonstrated a DFS encoded version of the key operation in one-way QIP: genuine teleportation in the form of a one-way information transfer protocol~\cite{2}. 
We demonstrate 
processed outcomes strikingly close to the ideal situation 
even when severe noise is present. Linear optics is at the forefront of experimental one-way QIP 
and represents an ideal testbed for 
our setup-independent model~\cite{n18}. Our results suggest that noise-resilient, large-scale quantum devices based on the one-way model are a realistic possibility.

\begin{figure}[t]
\psfig{figure=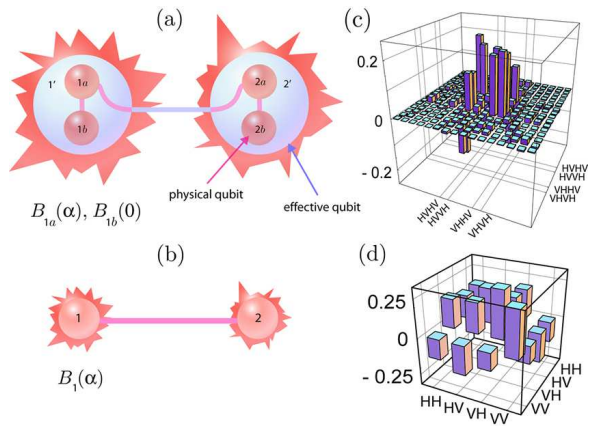,height=5.6cm}
\caption{In \textbf{(a)} (\textbf{(b)}) we show a DFS (standard) two-qubit cluster state. 
Effective qubits 
are represented by two physical qubits. \textbf{(c)} (\textbf{(d)}): Tomographic plots of the real parts of the experimental density matrices for four-qubit DFS (two-qubit standard) cluster state (imaginary parts on average $<0.05$).}
\label{fig1}
\end{figure}

{\it Model}.- For a standard cluster state, there are two types of single qubit measurements that enable a one-way quantum computer to operate~\cite{2}. First, by measuring the state of qubit $j$ in the computational basis $\{{\left| 0 \right\rangle}_{j},{{\left|1\right\rangle} }_{j}\}$, we disentangle it and shape the resource, leaving a smaller cluster. Second, in order to perform QIP, qubits must be measured in the basis $B_{j}(\alpha)=\{\mathop{{\left|\alpha_{\pm}\right\rangle}_{j}}=\frac{1}{\sqrt 2}({\left|0\right\rangle}\pm{e}^{i\alpha}{\left|1\right\rangle})_{j}\}$ ($\alpha\in\mathbb{R}$). This applies a single-qubit rotation $R_{z} (\alpha )={\rm e}^{-\frac{i}{2}\alpha\sigma_{z}}$, followed by a Hadamard operation $H$ to a logical qubit residing on site $j$ in the cluster ($\sigma_{x,y,z}$ are the Pauli matrices). With proper choices for the $B_{j}(\alpha )$'s, any quantum gate can be performed on a large enough cluster state. The structure and operation of a DFS cluster state are significantly different from the standard one: Each physical qubit $j$ initially prepared in ${\left| \pm \right\rangle}={\left|(\alpha =0)_{\pm}\right\rangle}$ in the standard model is replaced by a qubit-pair initially prepared in ${\left| \psi ^{-}\right\rangle}_{j'}=\frac{1}{\sqrt{2}}({\left|01\right\rangle}-{\left|10\right\rangle})_{j_{a}j_{b}}$. The encoding ${\left| 0_{E}  \right\rangle} \to {\left| 01 \right\rangle}$ and ${\left| 1_{E}  \right\rangle} \to -{\left| 10 \right\rangle}$ is used in order to pair two {\it physical} qubits into an {\it effective} one $j'$.
Next, instead of the standard controlled-phase gates 
applied to nearest-neighbor qubits for cluster state creation, 
now only qubits $j_a$ of nearest-neighbor $\ket{\psi^-}_{j'}$ pairs are involved (see Fig.~\ref{fig1} {\bf (a)} for two effective qubits $1'$ and $2'$). The resulting entangled resource has an effective structure exactly the same as a cluster state~\cite{n18}. Thus, one can use it for one-way QC, with the added benefit that it provides DFS protection from PD decoherence (described below). 
To perform DFS one-way processing, two single-qubit measurements are made on a pair for every corresponding qubit in the standard model that they replace (two-qubit entangled measurements can also be used~\cite{n18}). Measuring qubit $j_b$ in $B_{j_b}(0)$ removes the redundancy of the encoding~\cite{27}, qubit $j_a$ is then measured in $B_{j_a}(\alpha)$ as usual (for the final read-out of a computation, qubit $j_a$ will be measured in the computational basis). No decoding techniques are required, except for feedforward operations, as in the standard case~\cite{2,n18}. Our use of $\{ \ket{0_E},\ket{1_E} \}$ is known as {dual-rail} encoding~\cite{11}, enabling states within this subspace to be robust to various symmetrical multi-qubit 
randomized rotations~\cite{n18a}. The initial states $\ket{\psi^-}_{j'}$ used in the construction of a DFS cluster are {\it complete} DFS's in this respect. However, the entangling operations that put them into an effective cluster state limit their protection to strictly multiqubit PD noise~\cite{n18}, {\it i.e.} they allow a particular DFS encoding with respect to PD decoherence~\cite{11},  
which is a 
relevant source of decoherence in 
various quantum systems~\cite{24,n182}.
PD noise can be described as random phase shifts ${\left| k \right\rangle} \to e^{i\phi _{k} } {\left| k \right\rangle}$ on the state of a physical qubit. If $\phi_k$ is the same 
for every induced phase shift on two physical qubits encoding an {\it effective} one ({\it i.e.} the environment is symmetric), as the basis $\{ \ket{0_E},\ket{1_E} \}$ is invariant, so is the cluster state built out of it. 
Each pair of qubits ($j_a,j_b$) in a DFS cluster can even tolerate {\it different} symmetric PD noise (see Fig.~\ref{fig1} {\bf (a)}). 
Although our scheme is independent of the specific setting~\cite{n18} and is effective in any setup affected by symmetric PD noise, 
to demonstrate in a clear-cut way its performance 
under this noise, we use an all-optical setup. 
  
{\it Experimental implementation}.- We have used the scheme in Fig.~\ref{fig1} to demonstrate DFS quantum information transfer. A logical qubit ${\left|Q_{in}\right\rangle}=\mu{\left|0\right\rangle}+\nu{\left|1\right\rangle}$ (with $\left|\mu\right|^{2}+\left|\nu \right|^{2}=1$) is encoded on effective qubit $1'$ embodied by qubits $j_{a}=1a$ and $j_{b}=1b$. After the entangled resource is prepared, 
we obtain the DFS state ${\left|\Phi_{DFS}\right\rangle}=\mu{\left| 0_{E},+_{E}\right\rangle}_{1'2'}+\nu{\left|1_{E},-_{E}\right\rangle}_{1'2'}$. Information is transferred across $\ket{\Phi_{DFS}}$ by measuring the state of the qubits in $B_{1a}(\alpha)$ and $B_{1b}(0)$, where $\alpha$ determines the operation $HR_{z}(\alpha)$ on ${\left|Q_{in}\right\rangle}$. This can be compared to the case of a standard cluster (see Fig.~\ref{fig1} {\bf (b)} with $B_{1}(\alpha)$). 
\begin{figure*}[t]
\psfig{figure=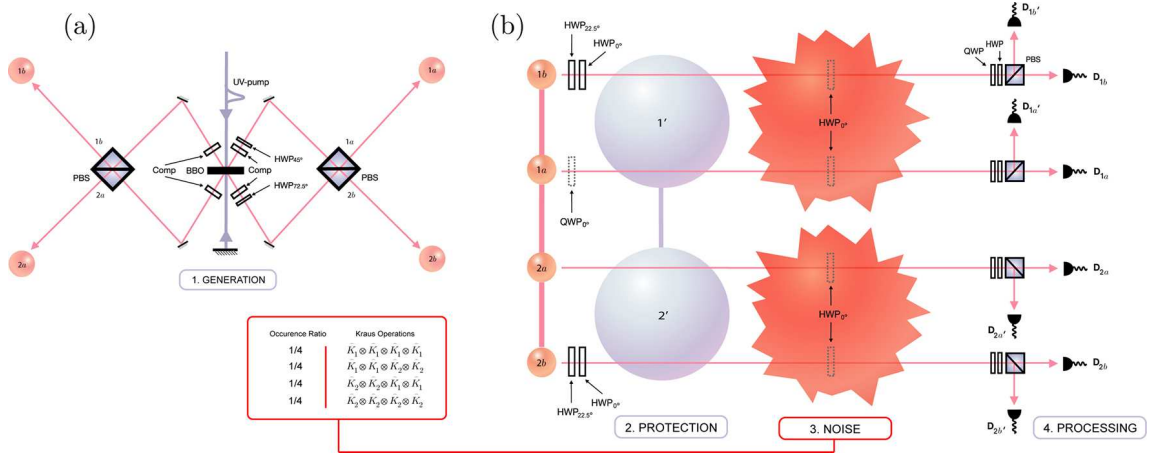,height=6cm}
\caption{\textbf{(a)}: The {\it generation} stage. \textbf{(b)}: The {\it protection}, {\it noise} and {\it processing} stages. A UV femto-second laser pulse with $1$ W of cw-power pumps a non-linear crystal (BBO) in a double-pass configuration. Compensation of walk-off effects in the crystal leads to the emission of Bell states (${\left| \Phi ^{-}  \right\rangle} $  and  ${\left| \Phi ^{+}  \right\rangle} $ in the forward and backward direction, respectively). Combination of these states on polarizing beam-splitters (PBS) and postselection yields a cluster state with a rate of $\sim1$ Hz in output modes $1a,b$ and $2a,b$. This is rotated into a DFS cluster by $\sigma _{x} \sigma _{z}$ operations on modes $1b$ and $2b$ using HWP's. Probe state ${\left| L \right\rangle} $ is encoded with an additional QWP in mode $1a$.  PD noise is implemented 
by inserting HWPs at $0^\circ$ ($\sigma_z$ operation) between the {protection} and {processing} stages. Polarization measurements in arbitrary bases are performed with  
PBS's, HWP's and QWP's.}
\label{fig2}
\end{figure*}
To test the DFS protection, PD is applied to the photonic qubits during information transfer in both the standard and DFS cases. For the latter (former), we consider symmetric noise on pairs $1'=(1a,1b)$ and $2'=(2a,2b)$ ($1$ and $2$).
We use the formalism of Kraus operators, useful in assessing noisy quantum systems~\cite{1}. In this context, a {\it channel} ({\it i.e.} a trace-preserving completely positive map) $\varepsilon$ transforms a state $\rho$ into $\varepsilon(\rho)=\sum_{i}\mathop{\hat{K}}\nolimits_{i}\rho\mathop{\hat{K}}\nolimits_{i}^{\dag}$ with the Kraus operators $\hat{K}_{i}$ such that $\sum _{i} \, \mathop{\hat{K}}\nolimits_{i}^{\dag}\mathop{\hat{K}}\nolimits_{i} =\openone$. For single-qubit PD, the non-zero Kraus operators are $\hat{K}_{i}=\sqrt{[1-(-1)^ie^{-\Gamma{t}}]/2}\sigma^{i-1}_{z}~(i=1,2)$
with $\Gamma$ the strength of the system-environment coupling and $t$ the interaction time. In our experiment, we simulate the case of 
$\Gamma{t}\!\to\!\infty$ (so that $\hat{K}_1=\openone,\,\hat{K}_2=\sigma_z$) corresponding to a full destruction of coherences in a single-qubit state~\cite{density}. In this limit, the noise experienced by each ($a,b$) pair is the same (but still {\it independent}). The noise can be different in general, making our DFS encoding suitable for other setups such as optical lattices and ion traps~\cite{n18}.
By means of quantum process tomography (QPT)~\cite{14,n18} we experimentally determine the {\it logical} transfer channels $\varepsilon_{\rm L} (\rho_{in})$  (with $\rho_{in}= \ketbra{Q_{in}}{Q_{in}}$) while the physical qubits are exposed to noise. With this apparatus, the reconstruction of the effect of $\varepsilon_{\rm L}$ on the set of logical input states ({\it probe states}) $\left\{{\left| 0 \right\rangle} ,{\left| 1 \right\rangle} ,{\left| + \right\rangle} ,{\left| L \right\rangle} \right\}$ is enough for a full characterization of the physical process encompassed by the logical channel~\cite{14}. Thus, we have implemented the DFS protocol by encoding the probe states onto photons $1a$ and $1b$ of the four-photon DFS cluster. Although arbitrary logical inputs 
are not possible in our setup, through tailored measurement patterns (as described below) it is possible to encode all probe states. 
We remark that in principle, arbitrary encoding is not necessary for sufficiently large clusters~\cite{2}.
To perform the information transfer protocol, we create a linear cluster state ${\left| \Phi _{lin}\right\rangle}$ by using the setup illustrated in Fig.~\ref{fig2} {\bf (a)}. This technique is a standard tool for the
generation of four-photon clusters~\cite{4, 7, 8} and ensures that photon loss and detector inefficiency do not affect the experimental results. ${\left| \Phi _{lin}  \right\rangle}$ is rotated into the DFS cluster $\left| \Phi _{DFS} \right\rangle=\frac{1}{2}(| 0101\rangle-|0110\rangle-| 1001\rangle-| 1010\rangle) _{1a1b2a2b}$ by applying $\sigma_{x}\sigma_{z}$ to qubits $1b$ and $2b$,
where ${\left|0\right\rangle}$ (${|1\rangle}$) is embodied by the horizontal (vertical) polarization of a photon. By means of over-complete state tomography~\cite{30}, we reconstruct the density matrix of the DFS cluster. We use 1296 measurements (each taking $350$ sec) and a maximum-likelihood (ML) function on all combinations of polarization-projections on the qubits, {\it i.e.} $\left\{{\left| 0/1 \right\rangle};{\left|\pm \right\rangle};{\left| L/R \right\rangle}\right\}$, where  ${\left|\pm\right\rangle}$ denote  $\pm45^\circ$ polarization and  ${\left| L/R \right\rangle}$ is for left/right circular polarization. The experimental 
state $\rho_{exp}$ has fidelity $F_{DFS}={\left\langle \Phi _{DFS}  \right|} \rho_{exp}^{DFS} {\left|\Phi_{DFS}\right\rangle}=0.70\pm 0.01$ ($F_{C}={\left\langle \Phi _{C}  \right|} \rho_{exp}^{C} {\left|\Phi_{C}\right\rangle}=0.74\pm 0.02$ with $\ket{\Phi_{C}}=(1/\sqrt{2})(\ket{0,+}+\ket{1,-})_{12}$) with the ideal DFS (standard) cluster state~\cite{uncertainties}. The state $\rho_{exp}^{C}$ is obtained by projecting photons $1b$ and $2a$ from $\rho_{exp}^{lin}$ onto $\ket{+}$. Tomographic plots of the
DFS and standard resources are in Figs.~{\ref{fig1}} {\bf (c)} and {\bf (d)}.
For the DFS cluster, by measuring photon $1b$ in ${\left| 1 \right\rangle}$, we encode the probe state ${\left| 0 \right\rangle}$ on effective qubit $1'$. Then measuring photon $1a$ in ${\left| + \right\rangle}$ we transfer the logical state across the cluster to $2'$ (embodied by $2a,b$). Analogously the other probe states are encoded and transferred by the measurement patterns $B_{1a}(0),\left|0\right\rangle_{1b}$ for $\left|{1}\right\rangle$, $B_{1a}(0),B_{1b}(0)$ for $\left|+\right\rangle$ and $B_{1a}(0)R_{z}(\frac{\pi}{2}),B_{1b}(0)$ for $\left| L \right\rangle$ ($R_{z}(\frac{\pi}{2})$ is realized with a quarter-wave plate (QWP) at $0^{\circ}$ placed in mode $1a$). For each of these input states, we perform two-qubit state tomography of the output qubits $2a$ and $2b$. We repeat this in the presence of PD noise
by properly applying half-wave plates (HWP's) at $0^\circ$ (realizing 
$\hat{K}_2=\sigma _{z}$) to the photons. 
As the occurrence-ratio 
in time of $\hat{K}_i$'s is dictated by the 
state given in~\cite{density} and shown in the box of Fig.~\ref{fig2} {\bf (b)}, we apply the HWP combinations in sequence, each for $1/4$ of the duration of the 
tomography process.
Such a realization of full 
symmetric PD noise complements simpler phase-flip mechanisms used in previous characterisations of photonic DFS's~\cite{21}. From the data for the probe states (1728 measurements) and QPT, we reconstruct the {\it logical} Kraus operators ({$\hat{K}_{{\rm L},i}$'s}) of the channel $\varepsilon_{\rm L}$~\cite{n18}.
The effect on a logical input state can be visualised using the correspondence between single-qubit density matrices and Bloch vectors. Taking many instances of ${\left|Q_{in}\right\rangle}$ (each for a value of $\mu$) and representing their output states from the reconstructed channel in a three-dimensional space, we depict the deformation of the single-qubit Bloch sphere induced by the experimental logical transfer channel. Fig.~\ref{fig3} {\bf (a)} shows the results for a DFS channel with and without noise. The information-protection is striking: the Bloch sphere for the noise-affected DFS state is almost identical to the no-noise case.  
The state fidelity averaged over ${\mu}$, quantifying the closeness of the experimental channels, is found to be $0.991\pm 0.003$.

The benefit of the DFS-protocol must be evaluated with respect to the standard case. We have run an experiment where a two-qubit cluster state is used for the same protocol. As before, the input state is transferred across the cluster with and without PD applied. The chosen measurements are $B_{1a}(0),\left| 0 \right\rangle_{1b},\left| + \right\rangle_{2a}$ for the probe state $\left|0\right\rangle$, $B_{1a}(0),|1\rangle_{1b},|+\rangle_{2a}$ for $|1\rangle$, $B_{1a}(0),|+\rangle_{1b},|+\rangle_{2a}$ for $|+\rangle$ and $B_{1a}(0),|+\rangle_{1b},|+\rangle_{2a}$ for $|L\rangle$ with an extra QWP, as in the DFS case.
The output states are analyzed using single-qubit state tomography while the channels with/without noise are constructed using QPT. Fig.~\ref{fig3} {\bf (b)} shows the resulting deformation of the Bloch sphere. Clearly the standard noise-free channel is very close to the analogous DFS case. However, if the information is not shielded in the DFS, the effects of the environment are severe. The output states in the presence of noise suffer strong decoherence: The Bloch sphere is shrunk and coherences are almost completely lost. The output density matrix has average fidelity of $0.994\pm 0.002$ with $\openone/2$ (resulting from a full PD process). The information in the standard resource has been greatly deteriorated.

\begin{figure}[t]
\psfig{figure=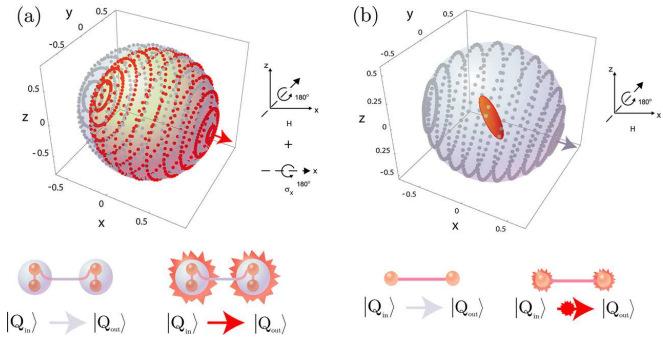,height=4.4cm}
\caption{\textbf{(a)} ({\bf (b)}): Experimental Bloch spheres for the channel using a DFS (standard) cluster state for outcomes  ${\left| +,+ \right\rangle} _{1a1b} $ (${\left| + \right\rangle} _{1}$). Underneath, we show the corresponding configuration for the protocol. The outer (inner) sphere shows the noise-free (noise-affected) case. 
The {\it quantum process fidelity} between the noise-free and noise-affected DFS (standard) protocol is $0.95 \pm 0.02$ ($0.53\pm{0.02}$, consistent with the value for a maximally mixed output state of a noise-affected standard resource)~\cite{14,n18,uncertainties,30}. The shape of the inner spheroid in \textbf{(a)} is due to small coherences in the reconstructed density matrix stemming from noise affecting the encoding of ${\left| L \right\rangle} $. 
The orientation of the poles of the Bloch sphere corresponding to the  ${\left| 0 \right\rangle} $ logical input state, shown by an arrow in \textbf{(a)} ({\bf(b)}), agrees with the expected $\sigma _{x} H$  $\left(H\right)$ operation applied during the protocol. The dots represent states corresponding to the action of the channel. 
}
\label{fig3}
\end{figure}
{\it Remarks}.- We have demonstrated a novel strategy to protect a one-way quantum protocol from symmetric PD noise. The effectiveness of the model has been verified in a proof-of-principle linear-optics experiment which has revealed excellent shielding of the processed information. 
Together with the setup-independent nature of the model~\cite{n18}, we expect our scheme to have important practical applications in any physical situation where symmetric PD noise is dominant~\cite{24, n182}. Conceptually, the idea can be generalized to other forms of symmetric noise. With a three-qubit encoding, one can achieve protection from general multi-qubit noise~\cite{n18, 24}. The resource overheads could be bypassed using additional degrees of freedom within the same physical information carrier
~\cite{n8}. The extension to more general environments of physical systems will require the integration of DFS's with other information-protection techniques~\cite{18, n18b}.

We thank M. Aspelmeyer,  
T. Jennewein, J. Kofler and T. Paterek for discussions. 
We acknowledge support from EPSRC, DEL, Leverhulme Trust, QIPIRC, FWF, EC
under the Integrated Project Qubit Application 
and U.S. Army Research Funded DTO.


\begin{thebibliography}{99}
\bibitem{density} In the DFS-encoded case, the action of noise on the state $\rho _{{\rm DFS}} =\mathop{{\left| \Phi _{{\rm DFS}}  \right\rangle} }\nolimits_{} {\left\langle \Phi _{{\rm DFS}}\right|}$ is given by $\varepsilon(\rho_{\rm DFS})=\frac{1}{4}[\rho _{{\rm DFS}} +\sigma _{z}^{1a} \otimes \sigma _{z}^{1b} \otimes \openone^{2a,2b} \rho _{{\rm DFS}}\sigma _{z}^{1a} \otimes \sigma _{z}^{1b}\otimes \openone^{2a,2b}$ $+\openone^{1a,1b} \otimes \sigma _{z}^{2a} \otimes \sigma _{z}^{2b} \rho _{{\rm DFS}}\openone^{1a,1b} \otimes \sigma _{z}^{2a} \otimes \sigma _{z}^{2b}$ $+\sigma _{z}^{1a} \otimes \sigma _{z}^{1b} \otimes \sigma _{z}^{2a} \otimes \sigma _{z}^{2b} \rho _{{\rm DFS}}\sigma _{z}^{1a} \otimes \sigma _{z}^{1b} \otimes \sigma _{z}^{2a}\otimes \sigma _{z}^{2b}]$. For a standard two-qubit cluster state with density matrix $\rho _{C} $ it is $\varepsilon(\rho_{C})=\frac{1}{4}[\rho_{C}+\sigma_{z}^{1}\otimes \openone^{2}\rho _{C}\sigma_{z}^{1}\otimes \openone^{2}$ $+\openone^{1}\otimes\sigma_{z}^{2}\rho_{C}\openone^{1}\otimes\sigma_{z}^{2}+\sigma_{z}^{1}\otimes\sigma_{z}^{2}\rho_{C}\sigma_{z}^{1}\otimes\sigma_{z}^{2}]$.
\bibitem{uncertainties} The uncertainty in the state (process) fidelity is from a 100 (50) run Monte Carlo simulation of the tomography analysis with Poissonian noise added to the count statistics in each run.
\end{thebibliography}
\end{document}